\documentclass{ws-ijmpb}

\usepackage{graphicx,amsmath,amssymb}

\begin{document}

\title{Attenuation and amplification of the transient current in
  nanojunctions with time-varying gate potentials}
\author{Eduardo C. Cuansing}
\address{Institute of Mathematical Sciences and Physics, University of the
  Philippines, Los Ba\~{n}os, 4031 Laguna, Philippines\\
  eccuansing@up.edu.ph}

\maketitle

\begin{abstract}
  We study charge transport in a source-channel-drain system with a
  time-varying applied gate potential acting on the channel. We calculate both
  the current flowing from the source into the channel and out of the channel
  into the drain. The current is expressed in terms of nonequilibrium Green's
  functions. These nonequilibrium Green's functions can be determined from
  the steady-state Green's functions and the equilibrium Green's functions of
  the free leads. We find that the application of the gate potential can
  induce current to flow even when there is no source-drain bias potential.
  However, the direction of the current from the source and the current to
  the drain are opposite, thereby resulting in no net current flowing within
  the channel. When a source-drain bias potential is present, the net current
  flowing to the source and drain can either be attenuated or amplified
  depending on the sign of the applied gate potential. We also find that
  the response of the system to a dynamically changing gate potential is not
  instantaneous, i.e., a relaxation time has to pass before the current
  settles into a steady value. In particular, when the gate potential is in
  the form of a step function, the current first overshoots to a maximum
  value, oscillates, and then settles down to a steady-state value.

~

\end{abstract}

\section{Introduction}
\label{sec:intro}

The relentless pursuit of miniaturization in electronics will eventually lead
to devices consisting of just a few atoms and molecules, with charge carriers
limited to move within reduced dimensions.\cite{aviram1974,heath2003,tao2006}
Given the small system sizes involved in molecular devices, a full
quantum-mechanical treatment is necessary in order for us to correctly
understand and predict the behavior and properties of the system. Several
groups have experimentally created molecular junctions and measured the
transport properties of these devices,\cite{reichert2002,cui2001,kubatkin2003}
including the appearance of negative differential
resistance,\cite{guisinger2004} Coulomb blockade,\cite{park2002} and Kondo
resonances.\cite{park2002,liang2002} An important issue in molecular
electronics is to determine the response of the system to time-varying forces
and system parameters. Experiments have been done to measure the dynamic
current in a quantum dot system with a time-varying bias
potential,\cite{lai2009,lu2003} in the switching of the bond angle in
bistable atoms,\cite{sweetman2011} in the controlled emission of electrons
from a nanoscale tip,\cite{kruger2011} and in the application of a gate
potential to a Wigner solid on liquid helium.\cite{rees2016} These experiments
show that whenever a change occurs in the system, the response is not
instantaneous, there is an initial overshoot, and current oscillations persist
for some time before approaching a steady-state value. Theoretical approaches
to model the response of quantum systems to time-dependent changes include
time-dependent density functional
theory,\cite{runge1984,stefanucci2004,bushong2005,kurth2010,varga2011}
Floquet theory,\cite{moskalets2002,kohler2005,arrachea2006,zhu2016}
density matrix renormalization group method,\cite{branschadel2010}
quantum Master equation,\cite{harbola2006,moldoveanu2010} Kadanoff-Baym
equations,\cite{myohanen2008,myohanen2009} and nonequilibrium Green's
functions method.\cite{jauho1994,waintal2013,haug2008,rammer1997,stefanucci2013,diventra2008,zhu2005,maciejko2006} The nonequilibrium Green's functions
method, in particular, is highly suitable in quantum transport calculations
of devices that can be partitioned into a source-channel-drain configuration.

Nonequilibrium Green's functions are functions of two time variables. In the
steady-state regime, the system satisfies time-translation invariance and
thus, one of the time variables can be integrated out and the Fourier
transform of the Green's functions can be utilized to re-express the functions
in the energy domain.\cite{datta2005} For the full time-dependent behavior,
including during the transient regime where time-translation invariance is
no longer satisfied, the Green's functions must be expressed in either two
time variables, two energy variables, or a combination of one time and one
energy variables. Two-time nonequilibrium Green's functions have been used
to study the response of a quantum dot device to a pulsed source-drain bias
potential\cite{jauho1994,zhu2005,xing2010}, including an analytically exact
solution for Lorentzian linewidths,\cite{maciejko2006} and a nano-relay
where the coupling between parts of the device is varied in
time.\cite{cuansing2011} Double-energy nonequilibrium Green's functions have
been used to study the response of a carbon nanotube transistor to a
time-dependent gate potential,\cite{kienle2010} in resonant-tunneling
devices,\cite{anantram1995} and in photon-assisted transport.\cite{platero2004}
Nonequilibrium Green's functions expressed as functions of a combination of
one time and one energy variables have been used to study the thermopower in a
quantum dot device with a time-dependent gate voltage.\cite{crepieux2011}
Recently, an energy-resolved self-energies approximation coupled with an
auxiliary-mode expansion scheme has been developed to eliminate one of the
time-dependence of the two-time nonequilibrium Green's
functions.\cite{croy2009} This scheme has been used to calculate the
time-dependent current in quantum dot systems in the presence of
electron-electron\cite{dong2015} and electron-phonon\cite{ding2016}
interactions. In this paper, we determine the two-time nonequilibrium Green's
functions directly to investigate the dynamic response of a
source-channel-drain device to a time-varying gate potential. We do not
calculate the nonequilibrium Green's functions in the energy domain to avoid
transforming these functions back to the time domain. Such transformations
could entail unphysical Gibbs oscillations in time that must be carefully
removed.\cite{press2007}

This paper is organized as follows. In Sec.~\ref{sec:model}, we model a
source-channel-drain system with a channel containing a single site and a
time-varying gate potential is acting on that channel. Expressions for the
electron and energy current in terms of nonequilibrium Green's functions
are derived. In Sec.~\ref{sec:negf}, the nonequilibrium Green's functions are
determined in terms of the steady-state Green's functions and the
equilibrium Green's functions of the free leads. In Sec.~\ref{sec:results},
we show the results of our calculations for various forms of the gate
potential. The summary and conclusion are stated in Sec.~\ref{sec:summary}.

\section{Model}
\label{sec:model}

\begin{figure}[h!]
  \centering{\includegraphics[width=2.3in]{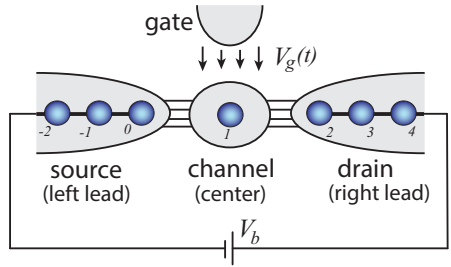}}
  \caption{Illustration of the source-channel-drain device with a single-site
    channel and a time-varying applied gate potential $V_g(t)$. The left and
    right leads are semi-infinite linear chains. $V_b$ is the source-drain
    bias potential. Sites are labeled consecutively.
  \label{fig:device}}
\end{figure}

Shown in Fig.~\ref{fig:device} is an illustration of the source-channel-drain
device. The channel consists of a single site, labeled $1$, where a
time-varying gate potential $V_g(t)$ is present. The left and right leads
are semi-infinite linear chains with the source as the left lead and the
drain as the right lead. The source-drain bias is a constant $V_b$.

We model the system using the tight-binding approximation. The total
Hamiltonian is $H = H^{\rm L} + H^{\rm R} + H^{\rm C} + H^{\rm LC} + H^{\rm RC}$.
The Hamiltonian for the left and right leads are
\begin{equation}
  \begin{split}
    H^{\rm L} & = \sum_k \varepsilon_k^{\rm L}\,a_k^{\dagger} a_k +
    \sum_{k<j} v_{kj}^{\rm L} \left( a_k^{\dagger} a_j + a_j^{\dagger} a_k\right),\\
    H^{\rm R} & = \sum_k \varepsilon_k^{\rm R}\,b_k^{\dagger} b_k +
    \sum_{k<j} v_{kj}^{\rm R} \left( b_k^{\dagger} b_j + b_j^{\dagger} b_k\right),
    \label{eq:Hleads}
  \end{split}
\end{equation}
where $a_k^{\dagger}$ ($a_k$) and $b_k^{\dagger}$ ($b_k$) are the creation
(annihilation) operators at site $k$ in the left and right leads,
respectively, the $\varepsilon_k^{\rm L}$ and $\varepsilon_k^{\rm R}$ are the
on-site energies at site $k$, the $v_{kj}^{\rm L}$ and $v_{kj}^{\rm R}$ are the
hopping parameters for nearest-neighbor sites $k$ and $j$, and the sums are
over all the sites in the leads. The Hamiltonian for the channel can be
separated into stationary and time-varying parts, i.e.,
$H^{\rm C} = H_0^{\rm C} + H^{\rm C}(t)$, where
$H_0^{\rm C} = \varepsilon_1^{\rm C}\,c_1^{\dagger} c_1$ and
$H^{\rm C}(t) = U(t)\,c_1^{\dagger} c_1$, where $c_1^{\dagger}$ and $c_1$ are
the creation and annihilation operators at site $1$ in the channel, the
real-valued $U(t) = -q V_g(t)$ is due to the time-varying potential that the
gate exerts on the site in the channel, and $q$ is the electron charge. The
Hamiltonian for the couplings between the leads and the channel are
\begin{eqnarray}
  H^{\rm LC} & = v_{01}^{\rm LC} \left( a_0^{\dagger} c_1 + c_1^{\dagger}
  a_0\right),\\
  H^{\rm RC} & = v_{21}^{\rm RC} \left( c_1^{\dagger} b_2 + b_2^{\dagger} c_1\right),
  \label{eq:Hcoupling}
\end{eqnarray}
where the coupling parameters
$v^{\rm LC} = v^{\rm CL}$ and $v^{\rm RC} = v^{\rm CR}$ are symmetric. The current
flowing out of the left lead can be determined from the changes in the
number operator, $N^{\rm L} = \sum_k a_k^{\dagger} a_k$, of the left lead:
\begin{equation}
  I^{\rm L}(t) = \left< -q \frac{d N^{\rm L}}{dt}\right> = -\frac{i q}{\hbar}
  \left< [H,N^{\rm L}]\right>
  = 2 q\,{\rm Re}\!\left[v_{01}^{\rm LC}\,G_{10}^{{\rm CL},<}(t,t)\right],
  \label{eq:Ileft}
\end{equation}
where the negative sign in the first equality implies the sign of $I^{\rm L}$
is positive if the current is flowing out of the left lead and
${\rm Re}[\,]$ means taking the real part. The lesser nonequilibrium Green's
function is defined as
$G_{10}^{{\rm CL},<}(t_1,t_2) = \frac{i}{\hbar} \langle a_0^{\dagger}(t_2)
c_1(t_1)\rangle$. The third equality in Eq.~(\ref{eq:Ileft}) is derived using
the fermion anti-commutation relation
$\{a_j,a_k^{\dagger}\} = \delta_{jk}$. Similarly, the current flowing into the
right lead is
\begin{equation}
  I^{\rm R}(t) = \left\langle q \frac{d N^{\rm R}}{dt}\right\rangle
  = -2 q\,{\rm Re}\!
  \left[ v_{21}^{\rm RC}\,G_{12}^{{\rm CR},<}(t,t)\right],
  \label{eq:Iright}
\end{equation}
where $G_{12}^{{\rm CR},<}(t_1,t_2) = \frac{i}{\hbar} \langle
b_2^{\dagger}(t_2) c_1(t_1)\rangle$ and the sign of $I^{\rm R}$ is positive if
current is flowing into the right lead. For both the left and right leads,
therefore, the sign of the current is positive whenever it is flowing from
the left to the right.

Another important dynamical variable is the amount of energy that the
electrons carry across the device, i.e., the heat due to electron flow.
This can be calculated from the rate of change of the lead Hamiltonian.
The heat resulting from the electrons flowing out of the left lead is
\begin{equation}
  Q^{\rm L}(t) = -\left\langle \frac{d H^{\rm L}}{dt}\right\rangle
  = 2\,{\rm Re}\!\left[ \left(\varepsilon_0^{\rm L} v_{01}^{\rm LC}
  + v_{00}^{\rm L} v_{01}^{\rm LC}\right) G_{10}^{{\rm CL},<}(t,t)\right]
  \label{eq:Qleft}
\end{equation}
while the heat due to electrons flowing into the right lead is
\begin{equation}
  Q^{\rm R}(t) = \left\langle \frac{d H^{\rm R}}{dt}\right\rangle =
    -2\,{\rm Re}\!\left[ \left(\varepsilon_2^{\rm R} v_{21}^{\rm RC}
    + v_{22}^{\rm R} v_{21}^{\rm RC}\right) G_{12}^{{\rm CR},<}(t,t)\right].
  \label{eq:Qright}
\end{equation}
Notice from the above expressions that the dynamics of the heat flowing
into or out of the leads simply follow, up to constant factors, the
dynamics of the flow of electrons into or out of those leads. Note, however,
that the dynamics considered here is due solely to the time-varying gate
potential. If the couplings between the leads and the channel are also
varied, heat flow may not follow the dynamics of the flow of the
electronic current.\cite{cuansing10,ludovico14} Furthermore, the heat
resulting from the flow of phonons and the heat due to electron-phonon
interactions are not incorporated in Eqs.~(\ref{eq:Qleft}) and
(\ref{eq:Qright}).

\section{Nonequilibrium Green's Functions}
\label{sec:negf}

\begin{figure}[h!]
  \centering{\includegraphics[width=2.8in]{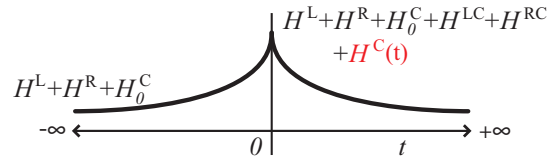}}
  \caption{Illustration of how components of the total Hamiltonian are
    switched on in time. 
    \label{fig:switchon}}
\end{figure}

The Green's functions are calculated based on the switch-on process described
in Fig.~\ref{fig:switchon}. At time $t \rightarrow -\infty$ the leads and the
channel are uncoupled. The leads are considered to be separately in
equilibrium at their respective temperatures and chemical potentials at this
time. The couplings are then adiabatically switched on such that at time
$t = 0$ they have reached their full values. The system is thus at a steady
state at this time. The gate potential is then switched on at time $t = 0$.

\begin{figure}[h!]
  \centering{\includegraphics[width=1.5in]{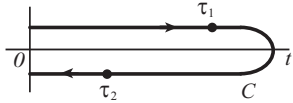}}
  \caption{Illustration of the Keldysh contour for time $t$. $\tau_1$ and
    $\tau_2$ are contour time variables and the arrows show the direction of
    contour ordering.
    \label{fig:contour}}
\end{figure}

The lesser nonequilibrium Green's functions can be determined from the
Keldysh formalism where time is extended into the complex
plane.\cite{jauho1994,rammer1997,stefanucci2013,diventra2008} We first define
the contour-ordered Green's function
$G_{10}^{\rm CL}(\tau_1,\tau_2) = -\frac{i}{\hbar}\langle {\rm T}_c\,
c_1(\tau_1) a_0^{\dagger}(\tau_2)\rangle$, where ${\rm T}_c$ is the
contour-ordering operator and $\tau_1$ and $\tau_2$ are contour time
variables on the Keldysh contour, as shown in Fig.~\ref{fig:contour}. To
calculate the Green's function we separate the total Hamiltonian into a
stationary and a time-varying part, $H = H_0 + H^{\rm C}(t)$, and the
transformation from the Heisenberg picture to the Interaction picture gives
\begin{equation}
  G_{10}^{\rm CL}(\tau_1,\tau_2) = -\frac{i}{\hbar}\left<{\rm T}_c\,
  e^{-\frac{i}{\hbar} \int_C H^{\rm C}(\tau') d\tau'} c_1(\tau_1) a_0^{\dagger}(\tau_2)
  \right>_0
  \label{eq:GCL}
\end{equation}
where the $\langle \ldots \rangle_0$ implies an ensemble average taken with
respect to the steady state. This Green's function can be expanded
perturbatively using diagrammatic techniques.

\begin{figure}[h!]
  \centering{\includegraphics[width=3.2in]{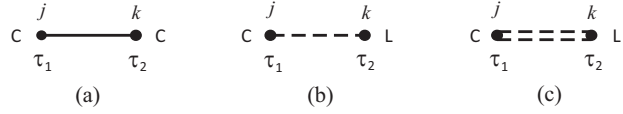}}
  \caption{Diagram representations of contour-ordered Green's functions. The
    diagrams represent (a) $G_{jk,0}^{\rm CC}(\tau_1,\tau_2)$, (b)
    $G_{jk,0}^{\rm CL}(\tau_1,\tau_2)$, and (c) $G_{jk}^{\rm CL}(\tau_1,\tau_2)$.
    The subscript $0$ indicates a steady-state Green's function.
    \label{fig:greenfuncs}}
\end{figure}

\begin{figure}[h!]
  \centering{\includegraphics[width=3.4in]{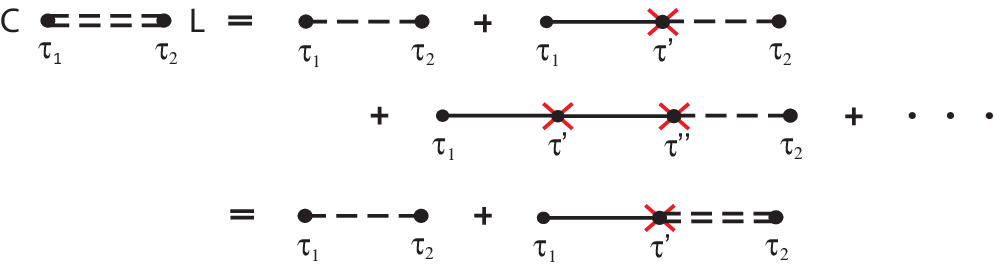}}
  \caption{The perturbation expansion of $G_{jk}^{\rm CL}(\tau_1,\tau_2)$ in
    terms of the steady-state Green's functions. Each vertex (a crossed dot)
    represents a $U(\tau)$.
    \label{fig:expansion}}
\end{figure}

Shown in Fig.~\ref{fig:greenfuncs} are the diagram representations of the
contour-ordered Green's functions needed for the expansion of
Eq.~(\ref{eq:GCL}). And shown in Fig.~\ref{fig:expansion} is the diagrammatic
expansion for $G_{10}^{\rm CL}(\tau_1,\tau_2)$.

From Fig.~\ref{fig:expansion}, the resulting iterative Dyson equation is
\begin{equation}
  G_{10}^{\rm CL}(\tau_1,\tau_2) = G_{10,0}^{\rm CL}(\tau_1,\tau_2)
  + \int_C d\tau'\,G_{11,0}^{\rm CC}(\tau_1,\tau')\, U_1(\tau')\,
  G_{10}^{\rm CL}(\tau',\tau_2),
  \label{eq:dyson}
\end{equation}
where the integral is along the Keldysh contour, the subscript $0$
indicates a steady-state Green's function, and the center steady-state Green's
function $G_{jk}^{\rm CC}(\tau_1,\tau_2) = -\frac{i}{\hbar}\langle {\rm T}_c\,
c_j(\tau_1) c_k^{\dagger}(\tau_2)\rangle$. To return the contour-time variables
back to the real-time axis, we apply analytic continuation and Langreth's
theorem.\cite{haug2008} The retarded and advanced Green's functions become
\begin{equation}
  G_{10}^{{\rm CL},\gamma}(t_1,t_2) = G_{10,0}^{{\rm CL},\gamma}(t_1-t_2)
  + \int_0^t dt'\,G_{11,0}^{{\rm CC},\gamma}(t_1-t')\, U_1(t')\,
  G_{10}^{{\rm CL},\gamma}(t',t_2),
  \label{eq:G10ra}
\end{equation}
where $\gamma = {\rm r}$ and ${\rm a}$, i.e., the retarded and advanced
Green's functions, respectively. For the lesser Green's function, following
Langreth's theorem
\begin{equation}
  \begin{split}
    G_{10}^{{\rm CL},<}(t_1,t_2) = ~& G_{10,0}^{{\rm CL},<}(t_1,t_2)
    + \int_0^t dt'\, G_{11,0}^{{\rm CC},r}(t_1,t')\, U_1(t')\,
    G_{10}^{{\rm CL},<}(t',t_2)\\
    & + \int_0^t dt'\, G_{11,0}^{{\rm CC},<}(t_1,t')\, U_1(t')\,
    G_{10}^{{\rm CL},a}(t',t_2).
  \end{split}
  \label{eq:G10less1}
\end{equation}
Notice that the second term in the right-hand side of the equation contains
$G_{10}^{{\rm CL},<}$. This can be solved by iteratively substituting
Eq.~(\ref{eq:G10less1}) whenever $G_{10}^{{\rm CL},<}$ appears in the
equation. For example, the first iteration is
\begin{equation}
  \begin{split}
    G_{10}^{{\rm CL},<}(t_1,t_2) & = G_{10,0}^{{\rm CL},<}(t_1,t_2)
    + \int_0^t dt'\, G_{11,0}^{{\rm CC},r}(t_1,t')\,U_1(t')\,
    G_{10,0}^{{\rm CL},<}(t',t_2)\\
    & + \int_0^t dt' \int_0^t dt''\,G_{11,0}^{{\rm CC},r}(t_1,t')\,
    U_1(t')\,G_{11,0}^{{\rm CC},r}(t',t'')\,U_1(t'')\,G_{10}^{{\rm CL},<}(t'',t_2)\\
    & + \int_0^t dt' \int_0^t dt''\,G_{11,0}^{{\rm CC},r}(t_1,t')\,U_1(t')\,
    G_{11,0}^{{\rm CC},<}(t',t'')\,U_1(t'')\,G_{10}^{{\rm CL},a}(t'',t_2)\\
    & + \int_0^t dt'\,G_{11,0}^{{\rm CC},<}(t_1,t')\,U_1(t')\,
    G_{10}^{{\rm CL},a}(t',t_2),
  \end{split}
  \label{eq:G10less2}
\end{equation}
where $G_{10}^{{\rm CL},<}$ appears in the third term of the right-hand side
of the equation. After we keep on iteratively substituting
Eq.~(\ref{eq:G10less1}) and combining terms, we get
\begin{equation}
  \begin{split}
    G_{10}^{{\rm CL},<}(t_1,t_2) & = G_{10,0}^{{\rm CL},<}(t_1,t_2)
    + \int_0^t dt'\, G_{11}^{{\rm CC},r}(t_1,t')\, U_1(t')\,
    G_{10,0}^{{\rm CL},<}(t',t_2)\\
    & + \int_0^t dt'\, G_{11,0}^{{\rm CC},<}(t_1,t')\, U_1(t')\,
    G_{10}^{{\rm CL},a}(t',t_2)\\
    & + \int_0^t dt' \int_0^t dt''\, G_{11}^{{\rm CC},r}(t_1,t')\, U_1(t')\,
    G_{11,0}^{{\rm CC},<}(t',t'')\, U_1(t'')\, G_{10}^{{\rm CL},a}(t'',t_2).
  \end{split}
  \label{eq:G10less}
\end{equation}
To determine $G_{10}^{{\rm CL},<}$ then the expressions for the center Green's
functions $G_{11}^{{\rm CC},<}$, $G_{11}^{{\rm CC},r}$, $G_{11}^{{\rm CC},a}$, and
their corresponding steady-state versions are needed. The nonequilibrium
${\rm CC}$ Green's functions can be determined by following the above same
procedure. The result is in the same form as Eq.~(\ref{eq:G10ra}) and
Eq.~(\ref{eq:G10less}) except for the replacement of all of the ${\rm L}$
superscripts with a ${\rm C}$ and all of the $0$ site labels with a $1$. In
addition, the same procedure can be employed to find the expression for
$G_{12}^{{\rm CR},<}$ for the right current in Eq.~(\ref{eq:Iright}). The
result is again of the same form as Eqs.~(\ref{eq:G10ra}) and
(\ref{eq:G10less}) except for the replacement of all of the ${\rm L}$
superscripts with an ${\rm R}$ and all of the $0$ site labels with a $2$.

The integrals in Eq.~(\ref{eq:G10less}) for $G_{10}^{{\rm CL},<}(t_1,t_2)$
require not just the steady-state Green's functions but also the
non-steady-state retarded and advanced Green's functions. We thus determine
the retarded and advanced Green's functions first. Notice that these Green's
functions are in iterative integral equations, as shown in Eq.(\ref{eq:G10ra}),
and can be solved following a numerical procedure. Equation~(\ref{eq:G10ra})
is in the form
\begin{equation}
  f(t_a,t_b) = f_0(t_a,t_b)
  + \int_{t_1}^{t_N} f_0(t_a,t') U(t') f(t',t_b) dt',
  \label{eq:generic}
\end{equation}
where $t_1 \leq t_a,t_b \leq t_N$. Replacing the integral with the
corresponding expression using numerical integration
\begin{equation}
  f(t_a,t_b) = f_0(t_a,t_b)
  + \Delta t \cdot \sum_{i=1}^N c_i f_0(t_a,t_i) U(t_i) f(t_i,t_b),
  \label{eq:numerical}
\end{equation}
where the $c_i$ are the numerical integration coefficients. A linear problem
can then be constructed in the form $A \vec{x} = \vec{b}$, where
\begin{equation}
  A = \left( \begin{array}{ccc}
    1-\Delta t~c_1 f_0(t_1,t_1)~U(t_1) & -\Delta t~c_2 f_0(t_1,t_2)~U(t_2)
    & \cdots\\
    -\Delta t~c_1 f_0(t_2,t_1)~U(t_1) & 1-\Delta t~c_2 f_0(t_2,t_2)~U(t_2)
    & \cdots\\
    \vdots & \vdots & \ddots \\
  \end{array} \right),
  \label{eq:A}
\end{equation}
is an $N \times N$ matrix,
\begin{equation}
  \vec{x} = \left( \begin{array}{c}
    f(t_1,t_b)\\
    f(t_2,t_b)\\
    \vdots\\
    f(t_N,t_b)\\
  \end{array} \right),~{\rm and}~
  \vec{b} = \left( \begin{array}{c}
    f_0(t_1,t_b)\\
    f_0(t_2,t_b)\\
    \vdots\\
    f_0(t_N,t_b)\\
  \end{array} \right),
  \label{eq:linearprob}
\end{equation}
are $N$-dimensional column vectors whose sizes depend on the length of the
time interval from $t_1$ to $t_N$ and the time step $\Delta t$. The
solution is calculated numerically from the inverse of $A$, i.e.,
$\vec{x} = A^{-1} \vec{b}$, using LU decomposition.\cite{press2007} Once
the retarded and advanced non-steady-state Green's functions are determined,
the integrals in Eq.~(\ref{eq:G10less}) can be calculated numerically using
the method of quadratures.\cite{press2007}

The steady-state Green's functions can be built from the adiabatic switch-on
of the coupling between the leads and the channel. Since time-translation
invariance is satisfied in the steady state, the Green's functions depend
only on the difference between its two time variables. The Fourier transforms
of these functions into the energy domain are therefore well-defined with
the resulting ${\rm CC}$ steady-state Green's functions as
\begin{align}
  G_{11,0}^{{\rm CC},r}(E) & = \left[(E + i \eta) - \varepsilon_1^{\rm C}
    - \Sigma_{11}^{{\rm C},r}(E)\right]^{-1},
  \label{eq:G110r}\\
  G_{11,0}^{{\rm CC},a}(E) & = \left( G_{11,0}^{{\rm CC},r}(E)\right)^\ast,
  \label{eq:G110a}\\
  G_{11,0}^{{\rm CC},<}(E) & = G_{11,0}^{{\rm CC},r}(E)\,\Sigma_{11}^{{\rm C},<}(E)\,
  G_{11,0}^{{\rm CC},a}(E),
  \label{eq:G110less}
\end{align}
where the self-energies are
\begin{equation}
  \Sigma_{11}^{{\rm C},\alpha} = v_{10}^{\rm CL}\,g_{00}^{{\rm L},\alpha}(E)\,
  v_{01}^{\rm LC} + v_{12}^{\rm CR}\,g_{22}^{{\rm R},\alpha}(E)\,v_{21}^{\rm RC},
  \label{eq:Sigma}
\end{equation}
where $\alpha = {\rm r}$, ${\rm a}$, and $<$, and the
$g_{00}^{{\rm L},\alpha}(E)$ and $g_{22}^{{\rm R},\alpha}(E)$ are the equilibrium
Green's functions for the left and right leads, respectively. The ${\rm CL}$
steady-state Green's functions are
\begin{align}
  G_{10,0}^{{\rm CL},r}(E) & =  G_{11,0}^{{\rm CC},r}(E)\,v_{10}^{\rm CL}\,
  g_{00}^{{\rm L},r}(E),\label{eq:G100r}\\
  G_{10,0}^{{\rm CL},a}(E) & = \left( G_{10,0}^{{\rm CL},r}(E)\right)^\ast,
  \label{eq:G100a}\\
  G_{10,0}^{{\rm CL},<}(E) & = G_{11,0}^{{\rm CC},r}(E)\,v_{10}^{\rm CL}\,
  g_{00}^{{\rm L},<}(E) + G_{11,0}^{{\rm CC},r}(E)\,\Sigma_{11}^{{\rm C},<}(E)\,
  G_{10,0}^{{\rm CL},a}(E).
  \label{eq:G100less}
\end{align}
The ${\rm CR}$ steady-state Green's functions have the same form as those
in Eqs.~(\ref{eq:G100r}) to (\ref{eq:G100less}) except for the replacement of
the ${\rm L}$ superscripts with ${\rm R}$ and the $0$ site labels with $2$.
Once the steady-state Green's functions in the energy domain have been
determined, their Fourier transforms into the time domain are required for
the calculation of the full time-dependent nonequilibrium Green's functions.

The equilibrium Green's functions for the left and right leads can be
determined from the solution of the equations of motion of the particles.
For the left lead,
\begin{align}
  g_{00}^{{\rm L},r}(E) & = 2 \frac{(E+i\eta)-\varepsilon_0^{\rm L}}{v^2}
  \pm 2 i \frac{\sqrt{v^2 - (\varepsilon_0^{\rm L}-E)^2}}{v^2},
  \label{eq:g00r}\\
  g_{00}^{{\rm L},a}(E) & = \left( g_{00}^{{\rm L},r}(E)\right)^\ast,
  \label{eq:g00a}\\
  g_{00}^{{\rm L},<}(E) & = - f^{\rm L}(E)\left( g_{00}^{{\rm L},r}(E)
  - g_{00}^{{\rm L},a}(E)\right),
  \label{eq:g00less}
\end{align}
where $v$ is the hopping parameter between site $0$ and site $-1$ in the
left lead and $f^{\rm L}(E) = [e^{(E-\mu_{\rm L})/k_B T_{\rm L}} + 1]^{-1}$ is the
Fermi-Dirac distribution function, where $\mu_{\rm L}$ is the chemical
potential and $T_{\rm L}$ is the temperature of the left lead, respectively.
The expressions for the equilibrium Green's functions for the right lead
are of the same form as Eqs.~(\ref{eq:g00r}) to (\ref{eq:g00less}).

To calculate, therefore, the left and right currents in Eqs.~(\ref{eq:Ileft})
and (\ref{eq:Iright}), respectively, the corresponding nonequilibrium Green's
functions $G_{10}^{{\rm CL},<}(t_1,t_2)$ and $G_{12}^{{\rm CR},<}(t_1,t_2)$ are
needed. These Green's functions can be expressed in terms of the steady-state
Green's functions, i.e., Eqs.~(\ref{eq:G10ra}) and (\ref{eq:G10less}), and,
in turn, the steady-state Green's functions can be expressed in terms of the
equilibrium Green's functions, i.e., Eqs.~(\ref{eq:G110r}) to
(\ref{eq:G100less}). The equilibrium Green's functions are calculated from
Eqs.~(\ref{eq:g00r}) to (\ref{eq:g00less}).

\section{Numerical Results}
\label{sec:results}

In our numerical calculations we use the following parameters:
$\varepsilon^{\rm L} = \varepsilon^{\rm R} = \varepsilon^{\rm C} = 0$ and
$v^{\rm L} = v^{\rm R} = v^{\rm LC} = v^{\rm RC} = 2~{\rm eV}$. For the left lead
$\mu_{\rm L} = E_F$ while for the right lead $\mu_{\rm R} = E_F - v_b$, where
we set the Fermi energy $E_F = 0$ and $v_b$ is the source-drain bias potential.
Both the left and right leads have the same temperature
$T_{\rm L} = T_{\rm R} = 300~{\rm K}$. We use a time step of
$\Delta t = 0.05~{\rm fs}$.

\begin{figure}[h!]
  \centering{\includegraphics[width=3in,clip]{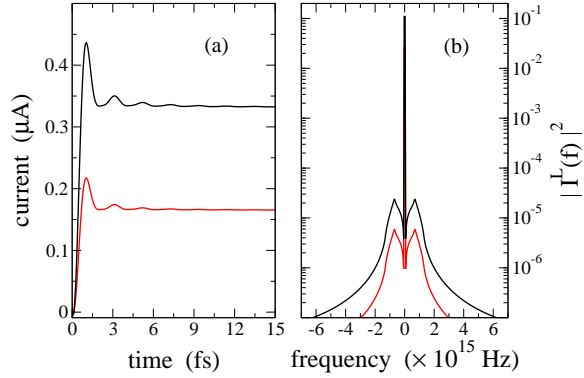}}
  \caption{(a) The current from the left lead $I^{\rm L}(t)$ as a function of
    time when the gate potential is switched-on as a step function at time
    $t = 0$. For the time interval up to $15~{\rm fs}$ and a time step of
    $\Delta t = 0.05~{\rm fs}$, the corresponding matrix $A$ in
    Eq.~(\ref{eq:A}) has size $N^2 = 90000$. (b) Semi-log plot of
    $\left| I^{\rm L}(f)\right|^2$ as a function of the frequency $f$ where
    $I^{\rm L}(f)$ is the Fourier transform of the current $I^{\rm L}(t)$. For
    both figures, the gate potentials are $0.1~{\rm eV}$ (dark line) and
    $0.05~{\rm eV}$ (light line, red online). There is no source-drain bias
    potential and both leads have the same temperature $T = 300~{\rm K}$.
  \label{fig:step}}
\end{figure}

Shown in Fig.~\ref{fig:step} are plots of the current flowing out of the left
lead as functions of time, $I^{\rm L}(t)$, and frequency, $I^{\rm L}(f)$. The
$I^{\rm L}(f)$ is the Fourier transform of its time-domain counterpart and
are calculated using the FFTW software package.\cite{fftw} The gate potential
is switched on in the form of a Heaviside step function at time $t = 0$.
There is no source-drain bias potential across the device, i.e.,
$v_b = 0$. The current, therefore, should not flow without the action of
the gate potential. At the transient regime as the gate potential is switched
on, the current takes time to react, overshoots to a large value, and then
oscillates down to a steady value. The height of the overshoot and the
amplitude of the oscillations depend on the strength of the gate potential.
In the plot of $|I^{\rm L}(f)|^2$ as a function of the frequency shown in
Fig.~\ref{fig:step}(b), there is a sharp peak around $f = 0$ indicating the
DC component of the current. There are two other peaks showing up at
$\pm~0.7 \times 10^{15}~{\rm Hz}$. These peaks correspond to the frequency
of oscillation of the current as it settles down from the transient to the
steady-state regime.

\begin{figure}[h!]
  \centering{\includegraphics[width=3in,clip]{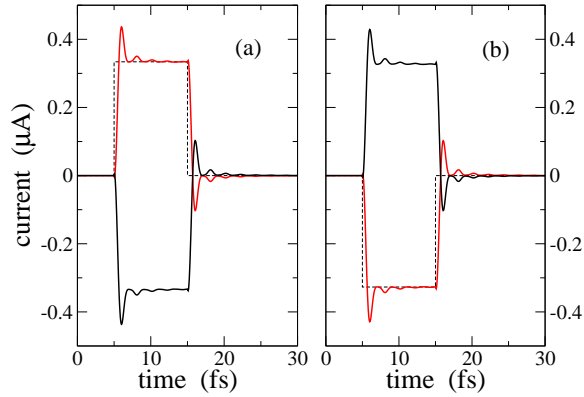}}
  \caption{The current from the left lead $I^{\rm L}(t)$ (light line, red
    online) and right lead $I^{\rm R}(t)$ (dark line) when a pulsed gate
    potential (represented by the dash line) acts on the channel. The gate
    potential in (a) is $0.1~{\rm eV}$ while in (b) it is $-0.1~{\rm eV}$
    during the pulse. There is no source-drain bias potential. The
    corresponding matrix $A$ has size $N^2 = 360000$.
  \label{fig:pulsenobias}}
\end{figure}

Shown in Fig.~\ref{fig:pulsenobias} are the plots of the left and right
current for a pulsed gate potential (represented by the dash lines) and
there is no source-drain bias potential. The pulse is on between
$t = 5~{\rm fs}$ and $t = 15~{\rm fs}$. Notice that when the gate potential
is $+0.1~{\rm eV}$ the direction of the left current is to the right while
the direction of the right current is to the left, as shown in
Fig.~\ref{fig:pulsenobias}(a). In contrast, when the gate potential has the
opposite sign, the directions of the currents are reversed, as shown in
Fig.~\ref{fig:pulsenobias}(b). Note that the charge carriers are electrons
and thus a positive gate potential is attractive to the particles while a
negative gate potential is repulsive. When the gate potential is positive,
therefore, electrons from both the left and right leads flow into the
channel. The opposite direction happens when the gate potential is negative. 
Notice too that the response of the current to a sudden change in the gate
potential is to overshoot, oscillate, and eventually settle down to a
steady value. The form of the current during the downward step pulse is a
mirror reflection of the form during the upward step pulse. Since there is
no applied source-drain bias potential, and therefore no preferred direction,
the symmetry in the dynamics of the current between the upward and downward
step pulses is expected. This symmetry is not satisfied when the linewidths
of the leads are in the wide-band limit\cite{jauho1994} or in a finite
interval\cite{zhu2005,maciejko2006} where not all energy levels in the leads
are available for the electrons to propagate through.

\begin{figure}[h!]
  \centering{\includegraphics[width=2.75in,clip]{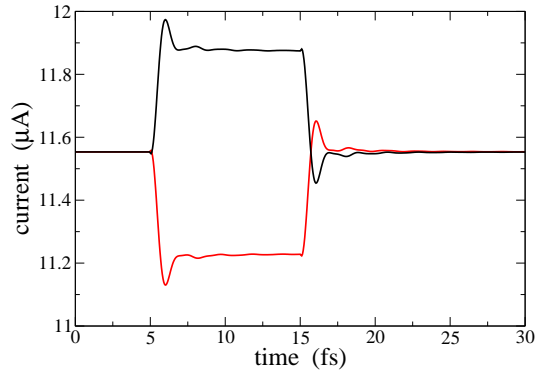}}
  \caption{The left current (light line, red online) and right current (dark
    line) when a pulsed $-0.1~{\rm eV}$ gate potential acts on the channel.
    The source-drain bias potential is $0.3~{\rm eV}$.
  \label{fig:pulsewithbias}}
\end{figure}

Since the effect of the gate potential is to induce current to flow in
opposite directions between the left and right leads, a gate potential
without an accompanying source-drain bias potential will not result in a
net flow of current. Shown in Fig.~\ref{fig:pulsewithbias} is the current
when a $0.3~{\rm eV}$ bias potential is applied acros the device and a
pulsed $-0.1~{\rm eV}$ gate potential acts on the channel. The pulse gate
potential is switched on at $t = 5~{\rm fs}$ and off at $t = 15~{\rm fs}$.
Both the left and right currents have the same steady values before the
pulse is switched on. Also, for long times after the pulse has been switched
off, both the left and right currents have the same steady values. These
steady values are consistent with the steady-state current values calculated
using the Landauer formula.\cite{datta1995} When the gate potential is on,
the current flowing out of the left lead is attenuated while that flowing
into the right lead is amplified.

\begin{figure}[h!]
  \centering{\includegraphics[width=3.4in,clip]{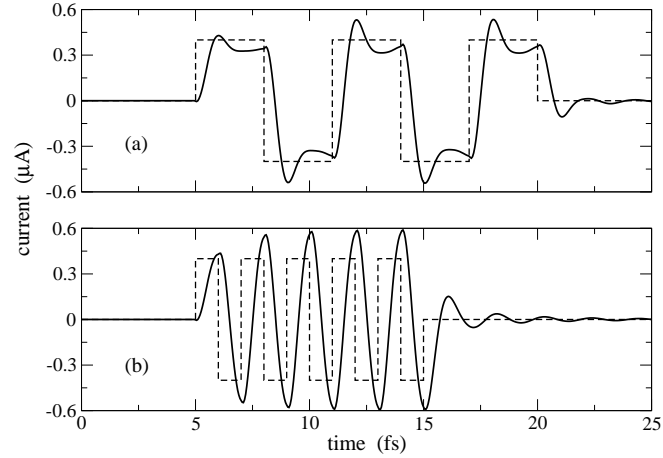}}
  \caption{The right current in a series of $0.3~{\rm eV}$ rectangular gate
    pulses (represented by the dash lines) of alternating signs with widths
    (a) $3~{\rm fs}$ and (b) $1~{\rm fs}$. There is no source-drain bias
    potential. 
  \label{fig:pulses}}
\end{figure}

Shown in Fig.~\ref{fig:pulses} is the current flowing into the right lead
when the gate potential is a series of rectangular pulses. The widths of the
pulses are $3~{\rm fs}$, shown in Fig.~\ref{fig:pulses}(a), and $1~{\rm fs}$,
shown in Fig.~\ref{fig:pulses}(b). There is no source-drain bias potential.
In the figures, the current overshoots whenever an upward or a downward
pulse occurs. In a $3~{\rm fs}$ pulse the width is long enough for the
current to make at least one oscillation after an overshoot. In the
$1~{\rm fs}$ pulse the width is too short for the current to complete an
oscillation. Also notice that the height of the overshoots increases for the
first few successive pulses. Because the gate potential is varying rapidly,
the current does not have enough time to reach a steady-state value and ends
up overshooting back and forth trying to catch up with the pulsing gate
potential.

\begin{figure}[h!]
  \centering{\includegraphics[width=3.4in,clip]{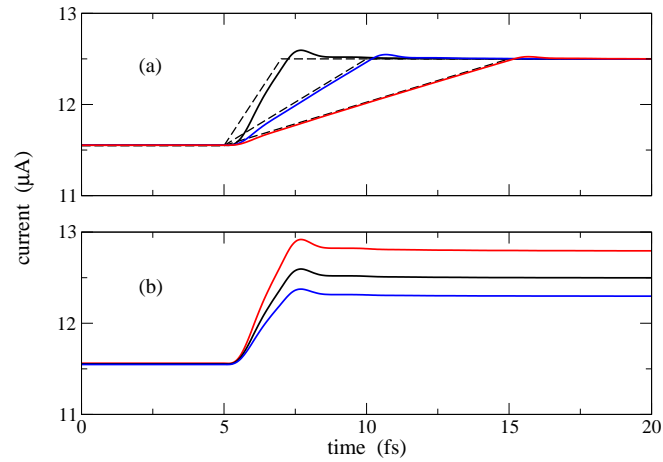}}
  \caption{The right current when the gate potential is switched on in the
    form of a ramp (represented by the dash lines). The gate potential is
    switched on at $t = 5~{\rm fs}$ and then gradually decreased to
    $-0.3~{\rm eV}$. In (a), the widths of the potential ramp are
    $2~{\rm fs}$ (dark line), $5~{\rm fs}$ (lighter line, blue online), and
    $10~{\rm fs}$ (light line, red online). In (b), the width of the ramp
    is $2~{\rm fs}$ while the leads-channel couplings are varied:
    $2~{\rm eV}$ (dark line), $2.2~{\rm eV}$ (lighter line, blue online),
    and $1.8~{\rm eV}$ (light line, red online). The source-drain bias
    potential is $0.3~{\rm eV}$.
  \label{fig:ramp}}
\end{figure}

The speed of the switch-on of the gate potential affects the height of the
current overshoot and the accompanying oscillations. Shown in
Fig.~\ref{fig:ramp} is the right current when the gate potential is switched
on gradually in the form of a ramp. In Fig.~\ref{fig:ramp}(a) the width of
the switch-on is varied from $2~{\rm fs}$ to $5~{\rm fs}$ and then to
$10~{\rm fs}$. For the $2~{\rm fs}$ ramp, the current rises behind the
gate potential, overshoots once the gate potential abruptly changes direction,
and the oscillates until it reaches a steady value. In the slightly softer
$5~{\rm fs}$ ramp, the current more closely follows the rising gate potential
but still overshoots, although with a shorter amplitude, when the gate
potential changes direction. For the soft $10~{\rm fs}$ ramp, the current
almost follows the rising gate potential and only softly overshoots when the
gate potential changes direction.

In Fig.~\ref{fig:ramp}(b) the width of the gate potential ramp is maintained
at $2~{\rm fs}$ and we vary the leads-channel coupling parameters from
$v^{\rm LC} = v^{\rm RC} = 1.8~{\rm eV}$ to $2~{\rm eV}$ and then to
$2.2~{\rm eV}$. Recall that for the leads, we use
$v^{\rm L} = v^{\rm R} = 2~{\rm eV}$. A different leads-channel coupling
parameter implies using a channel that is different from the leads. Notice
that one of the results of changing the leads-channel coupling is to vary the
steady-state value of the current. The initial line before the switch-on,
i.e., when $t<5~{\rm fs}$, are actually three lines that do not exactly have
the same values. The gate subsequently amplifies these differences as it is
switched on. We also see that a stronger leads-channel coupling would reduce
the height of the current overshoot as the gate potential changes direction. 

\section{Summary and Conclusion}
\label{sec:summary}

We model a source-channel-drain system with a channel containing a single
site. A time-varying gate potential is acting on the channel thereby making
the on-site energy of the site in the channel also vary in time. The electron
current flowing from the source (the left lead) and also the current flowing
into the drain (the right lead) can be expressed in terms of nonequilibrium
Green's functions. These nonequilibrium Green's functions can be expressed
in terms of the integrals of steady-state Green's functions. Furthermore, the
steady-state Green's functions can be written in terms of the integrals of the
equilibrium Green's functions in the leads. The equilibrium Green's functions
are calculated from the equation of motion of the free leads.

The response of the device to a change in the gate potential is not
instantaneous. A relaxation time has to pass before the current settles down
to a steady value. In the case when the gate potential is in the form of a
step function, the current overshoots to a maximum value, oscillates, and
then settles down to a steady-state value. The amplitude of the overshoot
and the oscillations depend on the strength and how fast the gate potential
is changing. The faster the gate potential is changing, the higher are the
resulting amplitudes in the transient current.

Electrons would flow when a time-varying gate potential is present even when
there is no source-drain bias potential. However, the direction of current
flow is opposite between the left and right leads, leading to no net current
flow in the channel. When a bias potential is present, the current flowing
across the channel can be attenuated or amplified depending on the sign of
the gate potential.

\section*{Acknowledgments}
The author is grateful to Jian-Sheng Wang, Gengchiau Albert Liang, Francis
Bayocboc, and Christian Laurio for insightful discussions. This work is
funded by the UP System Enhanced Creative Work and Research Grant
(ECWRG-2015-2-025).

\end{document}